\documentclass[3p,times,twocolumn]{article}

\usepackage{amssymb}

\usepackage[figuresright]{rotating}

\begin{document}
\begin{titlepage}


\begin{flushright} 
{ Preprint IFJPAN-IV-2016-28 
} 
\end{flushright}

\vskip 3 mm
\begin{center}

{ \bf\huge Tau lepton production and decays: perspective of multi-dimensional distributions and Monte Carlo methods }
\end{center}
\vskip 9 mm
\begin{center}
{\bf Z. Was}
\vskip 3 mm

{\em Institute of Nuclear Physics, Polish Academy of Sciences, PL-31342 Krakow}
\end{center}
\vskip 6 mm

\begin{abstract}

Status of    
  $\tau$ lepton decay Monte Carlo generator {\tt TAUOLA}, its main applications
and recent developments are reviewed.  It is underlined, that in 
recent efforts on development of new hadronic currents, the multi-dimensional 
nature of distributions of  the experimental data must be taken with a great care:
 lesson from  
comparison and fits to the BaBar and Belle
 data is recalled. It was found, that as in the past
 at a time of comparisons with  CLEO and ALEPH data,
proper fitting, to as detailed as possible representation of the 
experimental data, is essential for appropriate developments of 
models of $\tau$ decay dynamic. 

 This multi-dimensional nature of distributions is
also important  for observables where $\tau$ leptons are used to constrain 
experimental data.
In  later part of the presentation,
use of the {\tt TAUOLA} program for phenomenology of $W,Z,H$ decays at LHC 
is addressed, in particular in the context of the Higgs boson parity measurements.
Some new results, relevant  for QED lepton pair emission  are mentioned as well. 
\vskip 6 mm
\centerline{\bf Presented on the 14th International Workshop on Tau Lepton Physics}
\centerline{\bf 19-23 September, 2016, IHEP, Beijing, China } 
\end{abstract}

\vfill  
{\bf  IFJPAN-IV-2016-28, November 2016}





\end{titlepage}
\newpage
\section{Introduction}
\label{intro}

The {\tt TAUOLA} package
\cite{Jadach:1990mz,Jezabek:1991qp,Jadach:1993hs,Golonka:2003xt} for simulation
of $\tau$-lepton decays and
{\tt PHOTOS} \cite{Barberio:1990ms,Barberio:1994qi,Golonka:2005pn} for simulation of QED radiative corrections
in decays, are computing
projects with a rather long history. Written and maintained by
well-defined (main) authors, they nonetheless migrated into a wide range
of applications where they became ingredients of
complicated simulation chains. As a consequence, a large number of
different versions are presently in use. Those modifications, especially in case of
{\tt TAUOLA}, are   valuable from the physics point of view, even though they
 often did not find the place in the distributed versions of
the program.
From the algorithmic point of view, versions may
differ only in  details, but they incorporate many specific results from distinct
$\tau$-lepton measurements or phenomenological projects.
Such versions were mainly maintained (and will remain so)
by the experiments taking precision data on $\tau$ leptons.
Interesting from the physics point of view changes are still
developed in {\tt FORTRAN}.
That is why, for convenience of research partners, part of the
{\tt TAUOLA} need still to remain in {\tt FORTRAN} for a few forthcoming years.

Our presentation is organized as follows:
Section 2  is devoted to the discussion of initialization for 
 {\tt TAUOLA} hadronic currents. 
This point was already announced in  \cite{Was:2014zma} and presented in \cite{Chrzaszcz:2016fte}, that is why, we will address only those points which may be important for the future users.
In Section 3 we concentrate on   {\tt PHOTOS} Monte Carlo for
radiative corrections in decays. The new version of the program is  100 \%
in {\tt C++} and features emissions of light lepton pairs.
Section 4 is devoted to  applications of {\tt TAUOLA}
  for hard processes with final state $\tau$ leptons. In particular for
observable construction and evaluation of its sensitivity.
 The Neural Network techniques
in particular  Machine Learning (ML) techniques  were found to be useful.
 In this context we mention   {\tt TauSpinner} algorithm, which was found useful for evaluation 
of observable for Higgs boson parity measurement. We mention 
 other applications or tests; in particular in the domain of 
algorithm of calculating spin states of $\tau$ pairs in events where
high $p_T$ jets are present in $pp$ collisions.
 Summary Section 5, closes the presentation.
Because of the limited space of the contribution,
 some results  will not be presented in the
proceedings. They find their place in
publications, prepared with co-authors listed in the References.
For these works,  the present paper may serve as an advertisement.
\section{ Currents and structure of  {\tt TAUOLA} Monte Carlo}

The program structure did not change significantly 
since the previous
$\tau$ conference \cite{Was:2014zma}. Nowadays however, the {\tt C++ }
implementation become dominant for many aspects of  the project and that 
is why, the core part of 
the algorithm is gradually rewritten to introduce more modular structure and eventually to 
translate later the whole code into  {\tt C++} in quick well controlled steps.
The changes introduced so far, are  documented in \cite{Chrzaszcz:2016fte}.
This is more complicated task than it was for the completed already  transformation
of  {\tt PHOTOS} \cite{Davidson:2010ew}.
Constraints are  more complex.
Let us stress importance of the three aspects of the work:
(i)  construction and implementation of hadronic currents for $\tau$ decay 
currents obtained from models (inspired/evaluated from QCD) (ii) presentation of 
experimental data in a form suitable to fits (iii) preparations of algorithms 
and definition of distributions  useful for fits.

Already two years ago we have prepared two
new sets of currents; the first one based mainly on theoretical consideration, 
the second on an effort of BaBar collaboration. They are ready to be 
integrated into main distribution tar-balls for {\tt FORTRAN} and {\tt C++} 
applications, but we are still not sure if sufficient feed back is collected.
Weighted event 
techniques were found useful: for fits, and for evaluation of observables sensitivities
to model parameters. 

In description of $\tau$ decays and up to a precision level of about 0.2\% 
 hadronic currents   play the
dominant role in evaluation of systematic errors. At this precision level,
they constitute an important and well defined building 
block of the $\tau$-lepton decay description. It is  used 
since the first version of {\tt TAUOLA} Monte Carlo  generator as well. 
A multitude of models and parametrizations tuned to experimental data were 
used. It is at the same time source of the problems and  
opportunities  that precision 
of experimental data surpass significantly predictive power  of existing models.
Confrontation of model's predictions with experimental data with the help 
of multi-dimensional distribution has to be central for the future project developments.
This was pointed already in Ref.~\cite{Asner:1999kj} and its importance is clear from our recent 
experience 
of work on hadronic currents for $3\pi$ decay modes. Parametrization of 
Ref.~\cite{Shekhovtsova:2012ra}  had to be  modified in \cite{Nugent:2013hxa}, 
even though experimental input was 
enlarged with single one additional 1-dimensional distribution only. 
Note further comments in \cite{Chrzaszcz:2016fte}, explaining  potential limitations of such 
an approach.  
This is an essential topic which is behind recently changed 
re-organization of the {\tt TAUOLA} library.  
Appropriate choices should be coordinated between {\tt TAUOLA} Monte  Carlo and
fitting programs. Also, it should be possible, that hadronic currents 
can be coded in other than {\tt  FORTRAN} programing language. 
They have to be easy to modify by experimental user.

That is why, all variables of {\tt TAUOLA} initializations are stored now in 
{\tt COMMON} blocks, which can be accessed from other programming languages.
Version of Ref.~\cite{Chrzaszcz:2016fte} evolved from   
a variant presented on $\tau$ conference of 2004 \cite{Was:2004dg}
and used by the BaBar collaboration.
Program was supplemented with multitude of anomalous $\tau$ decay modes
as well as with parametrization of our theoretical works of the last decay 
(of different level of theoretical sophistication depending on decay channel). 
Details of decay matrix element parametrizations can be easily modified by the user, 
without the need of recompilation.

\section{{\tt PHOTOS} Monte Carlo for bremsstrahlung: \newline its systematic uncertainties}
\def\CCol{{\tt SANC}}
Over the last two years no major upgrades for functionalities 
were introduced into {\tt PHOTOS} Monte Carlo, except introduction 
of emission of lepton pairs. Documentation of the program~\cite{Davidson:2010ew}, was updated and
 published  finally.

Numerical tests for pair emission algorithm are advanced, but remain unpublished
as of the time when the proceedings material had to be submitted.

\section{  {\tt TAUOLA } - hard process - {\tt TauSpinner} algorithm }
In the development of packages such as {\tt TAUOLA} or {\tt PHOTOS}, questions
of tests and appropriate relations to users' applications are essential for
their
usefulness. In fact, user applications may be much larger in size and
human efforts than the programs discussed here.
Good example of such `user applications' are complete environments to simulate
physics process and control detector response at the same time.
Distributions of final state particles are not always of direct interest.
Often properties of intermediate states (manifesting through spin state of $\tau$-leptons):
e.g. coupling constants, masses or parity are of prime interest.
As a consequence, it is useful that such intermediate state properties are
under direct control of the experimental user and can be manipulated
to understand detector responses.

In that perspective, the algorithm of {\tt TauSpinner} \cite{Czyczula:2012ny} to study detector response to spin effects
in $Z, W$ and $H$ decays, represents a potentially  important development. 
The program is calculating weights corresponding to changes of the physics assumption. As an input,
events stored on the data file are used. There is no need to repeat simulations of the detector 
response whenever physics process is changed. The program has undergone 
several refinements \cite{Banerjee:2012ez,Przedzinski:2014pla},
and recently \cite{Kalinowski:2016qcd} where spin weight calculation 
taking into account matrix elements with two additional jets was introduced. 
This work, provided technical framework for  improvements in calculation 
of spin effects of Drell-Yan  $\tau$-lepton pair production process. It is straightforward to
extend  results of  \cite{Richter-Was:2016mal,Richter-Was:2016avq} to the case when spin 
effects are taken into account. Note, that electroweak corrections can be used
in calculation of complete spin correlations in $Z/\gamma^*$ mediated
processes. 

With the help of  {\tt TauSpinner} we could evaluate observable to study Higgs parity,
in its cascade decay with intermediate $\tau$ leptons \cite{Jozefowicz:2016kvz}. More precisely;
we have investigated the potential for measuring the CP state of the Higgs boson in $H \to \tau \tau$ decay, 
with consecutive  $\tau$-lepton decays in the channels: $\tau^{\pm} \to \rho^{\pm} \nu_{\tau}$ 
and $\tau^{\pm} \to a_{1}^{\pm} \nu_{\tau}$ combined. Subsequent decays 
$\rho^{\pm} \to \pi^{\pm} \pi^{0}$,  $ a_{1}^{\pm} \to \rho^{0} \pi^{\pm}$ and
 $ \rho^{0}\to \pi^{+}\pi^{-}$ were taken into account. 
We have extended 
  method of Ref.~\cite{Bower:2002zx}, where the acoplanarity angle for the planes build on 
the visible decay
products   $\pi^{\pm} \pi^0$ of
 $\tau^{\pm} \to \pi^{\pm} \pi^0 \nu_{\tau}$, were used. The 
angle  is sensitive to transverse spin correlations, thus to parity.

Also in the case of the cascade decays of $\tau \to a_1 \nu$, 
information on the CP state of Higgs can be extracted  
from the acoplanarity angles.
In the cascade decay 
$ a_{1}^{\pm} \to \rho^{0} \pi^{\pm},\; \rho^{0} \to \pi^+ \pi^-$ up to four planes 
can be defined,  thus  16 distinct acoplanarity angle distributions
 are available for
$H \to \tau \tau \to a_1^{+} a_1^{-} \nu \nu$. 
The distributions carry supplementary
but  correlated information. It is cumbersome to evaluate 
an overall sensitivity.

We have investigated the sensitivity potential  of such analysis, by developing and 
applying ML techniques. We 
quantified possible improvements when  multi-dimensional
phase-space of outgoing decay products directions is used, instead of
one-dimensional projections i.e. the acoplanarity angles. 

We have not taken into account ambiguities resulting from detector 
uncertainties or background contamination. We have concentrated on the 
usefulness of ML methods and $\tau \to 3\pi \nu$ decays for Higgs boson parity 
measurement. Let us quote an example of numerical result taken from Ref.~\cite{Jozefowicz:2016kvz} 
and recall the table \ref{tab:DeepLearn}.

\begin{table*}
  \begin{tabular}{lrrrr}
\hline
\hline
  Features/variables      & Decay mode: $\rho^{\pm}- \rho^{\mp}$    &  Decay mode: $a_1^{\pm} - \rho^{\mp}$   & Decay mode:  $a_1^{\pm} - a_1^{\mp}$  \\ 
                          & $\rho^{\pm} \to \pi^{0}\ \pi^{\pm}$   &  $ a_{1}^{\pm} \to \rho^{0} \pi^{\mp},\ \rho^{0} \to  \pi^{+} \pi^{-}$  
                                                               &  $ a_{1}^{\pm} \to \rho^{0} \pi^{\pm},\ \rho^{0} \to  \pi^{+} \pi^{-}$             \\ 
                          &                                   &  $\rho^{\mp} \to \pi^{0}\ \pi^{\mp}$   &                                         \\
  \hline
  True classification                             & 0.782       &  0.782          &  0.782    \\
  $\varphi^*_{i,k}$                                       & 0.500       &  0.500         &  0.500     \\
  $\varphi^*_{i,k}$ and $y_i, y_k$                             & 0.624       &  0.569          &  0.536    \\
    4-vectors                                                & 0.638       &  0.590          &  0.557    \\
  $\varphi^*_{i,k}$, 4-vectors                            & 0.638       &  0.594          &  0.573    \\
  $\varphi^*_{i,k}$, $y_i, y_k$ and $m^2_i, m^2_k$                   & 0.626       &  0.578          &  0.548    \\
  $\varphi_{i,k}^*$, $y_i$, $y_k$, $m^2_i$, $m^2_k$ and 4-vectors         & 0.639       &  0.596          &  0.573    \\
\hline
\hline
\end{tabular}
\caption{ Average probability $p_i$ 
(calculated as explained in  Ref.~\cite{Jozefowicz:2016kvz})
that a model predicts correctly event $x_i$ to be of a  type $A$ (scalar),
with training being performed for separation between type $A$ and $B$ (pseudo-scalar). 
\label{tab:DeepLearn}}
\end{table*}

\section{Summary and future possibilities}

Versions of the hadronic currents available for the {\tt TAUOLA} library
until recently, were often based on old models and experimental data of 90's.
An alternative implementation of  currents, based on the Resonance Chiral Lagrangian,
or other approaches is prepared
and tested for the decay channels to 2, 3, 4 and 5 pions but it was not 
confronted with the experimental distributions of multi dimensional nature. 
Also collecting feed 
back from community of users may not have been completed.
Parametrizations used for the default simulations in BaBar collaboration is now
available. Further  advantage of this new option is a
multitude of rare and anomalous $\tau$ decay channels. In fact, such options  were extended 
even further. With the help of {\tt C++} interface, user provided hadronic current(s) 
or decay matrix element(s),
can be used to replace the ones of the library, at any moment of program execution. 

Methods for   confrontation of program predictions
with the experimental data are not developing fast.
Solutions 
postulated already in \cite{Asner:1999kj} are still not used in full.
 We rely on 
comparison of results with one-dimensional histograms of invariant 
masses formed from sub-groups of $\tau$ decay products. They have to be  defined for each decay channel 
separately and unfolded from the experimental backgrounds.

The status of
associated projects: {\tt TAUOLA universal interface } and {\tt TauSpinner}
was reviewed. Also new results for 
the high-precision version of  {\tt PHOTOS} for QED radiative corrections in
decays, were mentioned. 
All programs are ready for {\tt C++} applications
thanks to the {\tt HepMC} interfaces.

Presentation of the {\tt TAUOLA} general-purpose {\tt C++} interface
was given and its applications were shown.  An algorithm for study, with the help of weights calculated from
kinematic of events stored on data files,  detector responses to spin effects 
and production process variants in
$Z$, $W$ and $H$ decays was shown. The corresponding program {\tt TauSpinner},
is useful, e.g. to study Higgs parity sensitive observables at LHC. 
In this case, efficiency  of multi-dimensional distributions to constrain Higgs parity in 
cascade decays staring from  $H \to \tau^+\tau^-$, $\tau^\pm \to \nu_\tau a_1^\pm$,
was demonstrated.

\vskip 2 mm
\centerline{\bf Acknowledgements}

{\small 
The work on {\tt TAUOLA} would not be possible without continuous help an encouragements
from experimental colleagues.
The work of all  co-authors of the papers devoted to  {\tt TAUOLA} development 
was of great importance.
I hope, that it is clearly visible from  my contribution. 
This research was supported in part by
the Research Executive Agency (REA) of the European Union under
the Grant Agreement PITN-GA-2012-316704 (HiggsTools) and by funds of Polish National Science
Centre under decision UMO-2014/15/ST2/00049.}









\end{document}